\documentclass[12pt]{iopart}
\usepackage{graphicx}
\expandafter\let\csname equation*\endcsname\relax
\expandafter\let\csname endequation*\endcsname\relax
\usepackage{mathtools}
\usepackage{amsmath,amssymb}

\usepackage[dvips]{color}
\usepackage{array}
\usepackage{stackrel}
\usepackage{tensor}

\usepackage{curves}
\usepackage[titletoc,title]{appendix}
\usepackage{cleveref}

\newcommand{\si}{\sigma}
\newcommand{\al}{\alpha}
\newcommand{\be}{\beta}

\newcommand{\ep}{\epsilon}
\newcommand{\pa}{\partial}
\newcommand{\da}{\dagger}
\newcommand{\ds}{\displaystyle}

\begin{document}

\title[The hidden symmetry of the asymmetric quantum Rabi model]
{The hidden symmetry of the asymmetric quantum Rabi model}

\author{Vladimir V. Mangazeev$^1$, Murray T. Batchelor$^{1,2,3}$ and Vladimir V. Bazhanov$^{1}$
}

\address{$^1$ Department of Theoretical Physics, Research School of Physics,
The Australian National University, Canberra ACT 2601, Australia}
\address{$^2$  Mathematical Sciences Institute,
The Australian National University, Canberra ACT 2601, Australia}
\address{$^3$ Centre for Modern Physics, Chongqing University, Chongqing 40444, China}

\begin{abstract}
The asymmetric quantum Rabi model (AQRM) exhibits level crossings in the eigenspectrum for the values $\ep \in \frac12 \mathbb{Z}$
of the bias parameter $\ep$.
Such level crossings are expected to be associated with some hidden symmetry of the model.
The origin of this hidden symmetry is established by finding the operators which commute with the AQRM hamiltonian at these special values.
The construction is given explicitly for the first several cases $\ep=0,\frac{1}{2},1$
and can be applied to other related light-matter interaction models for which similar
level crossings have been observed in the presence of a bias term.
\end{abstract}
\maketitle

A challenge in understanding fundamental models of light-matter interaction is to establish the origin of the observed hidden symmetry of the
asymmetric quantum Rabi model (AQRM).
The AQRM describes the interaction of a two-level system (the qubit) with a single mode bosonic field.
The asymmetry of the model is due to the applied bias term which breaks the well known $\mathbb{Z}_2$
symmetry of the quantum Rabi model \cite{Braak2019}.
In the absence of bias, the AQRM reduces to the quantum Rabi model, which, along with other related light-matter interaction models,
have a long history of inspiring developments from both the physical and mathematical perspectives \cite{BCBS}.
The bias term plays an important role in the realisation of the AQRM in quantum circuit devices \cite{review1,review2,review3}.
The $\mathbb{Z}_2$ symmetry of the quantum Rabi model is associated with level crossings in the eigenspectrum.
The effect of the bias term is to break these level crossings.
However, it was observed that level crossings in the eigenspectrum of the AQRM are also present for the values
$\ep \in \frac12 \mathbb{Z}$ of the bias parameter $\ep$ \cite{B2011}.
These crossing points are determined by constraint polynomials \cite{LB,W,M,KRW}.
More generally these polynomials fit within the general analytic solution obtained for the AQRM \cite{B2011,Chen,Lee,MPS,XZBL}.
The level crossings are physically important because they induce conical intersection points in the energy landscape \cite{BLZ},
for which the geometric phases associated with the Dirac-like cones have recently been calculated \cite{LFTB}.

The origin of the level crossings in the AQRM, and whether or not they are the result of some hidden symmetry, has remained a  puzzle.
A numerical study \cite{A} suggests that, unlike for the $\mathbb{Z}_2$ symmetry of the quantum Rabi model, there is no parameter-independent
partitioning of the Hilbert space, implying that any hidden symmetry operator must depend on the system parameters.
In earlier work \cite{Gardas} a connection is made with a Riccati equation, although no explicit solutions are known.

In this Letter we uncover the hidden symmetry of the AQRM by explicitly constructing
the operators which commute with the AQRM hamiltonian.
This is done on a case by case basis.

First, we search for such operators in the form of  $2\times2$ matrices in the qubit space with matrix
elements being
operators in the bosonic Hilbert space (Fock space). This results in a system of recurrence relations
for their coefficients in the Fock basis.
Imposing certain symmetry conditions we derive a system of PDEs for generating functions of these coefficients.
Then we rewrite these equations in symmetric variables and show that for $\ep \in \frac12 \mathbb{Z}$
there exists a nontrivial operator $J$ such that a product of $J$ and the parity
operator $\mathcal{P}$ is a local matrix operator in the bosonic space.
Therefore, $J^2$ is also a local matrix operator in the Fock space. For the first few values of $\ep \in \frac12 \mathbb{Z}_+$
we checked that $J^2$ is a polynomial in the hamiltonian of the AQRM of degree $2\epsilon$.
This naturally generalises the $\mathbb{Z}_2$ symmetry of the quantum Rabi model where $J^2 = 1$.
Our procedure can be applied to other related light-matter interaction models for which similar level crossings in the
eigenspectrum have been observed in the presence of a bias term \cite{LB2020}.

\newpage
The AQRM is defined by the hamiltonian
\begin{equation}
H=\omega a^\dagger a +g\si_x(a+a^\dagger)+\Delta \,\si_z+{\ep}\si_x,\label{eq1}
\end{equation}
where $\si_x$, $\si_z$ are the Pauli matrices for the two-level system with level splitting $2 \Delta$.
The operators $a$ and $a^\dagger$ satisfy the oscillator algebra
\begin{equation}
[a,a^\dagger]=1.\label{eq2}
\end{equation}
The interaction between the two systems is via the coupling $g$.
This hamiltonian acts in $\mathbb{C}^2\otimes\mathcal{H}$, where $\mathcal{H}$ is
the Fock space.
In the standard basis where  $\si_z$ is diagonal,
we can write the hamiltonian as a $2\times 2$ matrix with operator entries
\begin{equation}
H=\left(\begin{matrix}
a^\dagger a+\Delta & g(a+a^\dagger)+\ep \\
g(a+a^\dagger)+\ep & a^\dagger a-\Delta
\end{matrix}\right), \label{eq3}
\end{equation}
where we have set the frequency $\omega=1$.
A dependence on $\omega$ can be easily restored by rescaling all constants by $\omega$.

Under the adjoint action such that  $(a^\da)^\da=a$, the hamiltonian \eqref{eq3} is self-adjoint, $H^\da=H$.
It also has an obvious symmetry
\begin{equation}
\si_x H(\Delta) \si_x=H(-\Delta).\label{eq3a}
\end{equation}
By changing to the basis in $\mathbb{C}^2$ where $\si_x$ is diagonal,
 we can also bring the hamiltonian \eqref{eq3} to the form
\begin{equation}
\widetilde{H}=UHU^{-1}=\left(\begin{matrix}
a^\dagger a+g(a+a^\dagger)+\ep & \Delta \\
\Delta & a^\dagger a-g(a+a^\dagger)-\ep
\end{matrix}\right),\label{eq4b}
\end{equation}
where
\begin{equation}
U=\frac{1}{\sqrt{2}}\left(\begin{matrix}
1&1 \\
1&-1
\end{matrix}\right).\label{eq4a}
\end{equation}

Establishing the origin of the hidden symmetry of this model is equivalent to finding an operator ${J}$ which commutes with $H$
\begin{equation}
[J,H]=0. \label{eq5}
\end{equation}
Obviously any series of the form
\begin{equation}
\sum_{i=0}^M \al_i H^i \label{eq4}
\end{equation}
commutes with $H$ for any positive integer $M$. This is a ``gauge'' degree of freedom
which we will remove at the end.

We will search for the operator $J$ as a $2\times2$ matrix with operator entries in the Hilbert
space $\mathcal{H}$
\begin{equation}
J=\left(\begin{matrix}
\mathcal{A}(a^\da,a) & \mathcal{B}(a^\da,a) \\
\mathcal{C}(a^\da,a) & \mathcal{D}(a^\da,a)
\end{matrix}\right),\label{eq6}
\end{equation}
where
\begin{equation}
A(a^\da,a)=\sum_{m,n=0}^\infty a_{m,n}(a^\da)^ma^n,\quad
\mathcal{B}(a^\da,a)=\sum_{m,n=0}^\infty b_{m,n}(a^\da)^ma^n, \label{eq7}
\end{equation}
\begin{equation}
\mathcal{C}(a^\da,a)=\sum_{m,n=0}^\infty c_{m,n}(a^\da)^ma^n,\quad
\mathcal{D}(a^\da,a)=\sum_{m,n=0}^\infty d_{m,n}(a^\da)^ma^n,\label{eq8}
\end{equation}
with all coefficients depending on the parameters $\Delta$, $g$ and $\ep$.
We also imply that all operators are normally ordered, i.e., all operators $a^\da$ are always to
the left of $a$'s.

Substituting (\ref{eq6})-(\ref{eq8}) into \eqref{eq5} and using \eqref{eq2}
one can obtain the following set of recurrence relations for the coefficients
\begin{align}
&(m-n)a_{m,n}+{\ep}(c_{m,n}-b_{m,n})+\nonumber \\
&g(c_{m-1,n}+c_{m,n-1}+(m+1)c_{m+1,n}-b_{m,n-1}-b_{m-1,n}-(n+1)b_{m,n+1})=0,\nonumber\\
&(m-n)d_{m,n}+{\ep}(b_{m,n}-c_{m,n})+\nonumber \\
&g(b_{m-1,n}+b_{m,n-1}+(m+1)b_{m+1,n}-c_{m,n-1}-c_{m-1,n}-(n+1)c_{m,n+1})=0,\nonumber\\
&(m-n+2\Delta)b_{m,n}+{\ep}(d_{m,n}-a_{m,n})+\nonumber \\
&g(d_{m-1,n}+d_{m,n-1}+(m+1)d_{m+1,n}-a_{m,n-1}-a_{m-1,n}-(n+1)a_{m,n+1})=0,\nonumber\\
&(m-n-2\Delta)c_{m,n}+{\ep}(a_{m,n}-d_{m,n})+\nonumber \\
&g(a_{m-1,n}+a_{m,n-1}+(m+1)a_{m+1,n}-d_{m,n-1}-d_{m-1,n}-(n+1)d_{m,n+1})=0.\nonumber\\ \label{eq9}
\end{align}
Boundary conditions imply that all coefficients with negative indices are equal to zero.

In general, the system \eqref{eq9} has infinitely many solutions.
We can start excluding coefficients with smaller indices
in terms of coefficients with larger indices but this is a non-ending process with a growing number
of degrees of freedom. One can impose terminating conditions on all coefficients and search for
polynomial solutions. However, we found that such solutions always reduce to polynomials \eqref{eq4}
in the hamiltonian $H$. Therefore, we need to work with infinite series.

It is convenient to rewrite difference equations \eqref{eq9} for coefficients in the form of differential equations
for their generating functions. For the operator $\mathcal{A}(a^\da,a)$ we define the generating function
$A(x,y)$ by
\begin{equation}
A(a^\da,a)=\sum_{m,n=0}^\infty a_{m,n} \, (a^\da)^ma^n,\quad
A(x,y)=\sum_{m,n=0}^\infty a_{m,n} \, x^m y^n, \quad\mbox{etc}.\label{eq9c}
\end{equation}

We will further restrict
ourselves to the case of self-adjoint operators $J$,
\begin{equation}
J=J^\da \label{eq9a}
\end{equation}
with real coefficients $a_{m,n}$, $b_{m,n}$, $c_{m,n}$ and $d_{m,n}$ for real $\Delta$ and $g$.
This reality condition is natural and is consistent with further analysis.
The  condition \eqref{eq9a} implies the following symmetries of coefficients
\begin{equation}
a_{m,n}=a_{n,m},\quad c_{m,n}=b_{n,m},\quad d_{m,n}=d_{n,m}.\label{eq9b}
\end{equation}

In addition, we assume that $J$ satisfies the symmetry \eqref{eq3a}
\begin{equation}
\si_x  \, J(\Delta) \, \si_x =J(-\Delta).\label{eq12}
\end{equation}
We notice that by multiplying $J$ by $\Delta$ we can change the sign of $J(-\Delta)$ in the RHS of \eqref{eq12}.

To summarize, we made the following assumptions: the operator $J$  is self-adjoint;
it has real coefficients in the oscillator basis
and it satisfies \eqref{eq12}.

Taking into account \eqref{eq12}
we can write
\begin{align}
&A(x,y)=D_+(x,y)-\Delta D_-(x,y), \quad D(x,y)=D_+(x,y)+\Delta D_-(x,y),\nonumber\\
&B(x,y)=B_+(x,y)+\Delta B_-(x,y),\quad C(x,y)=B_+(x,y)-\Delta B_-(x,y),\label{eq13}
\end{align}
where  $D_\pm(x,y)$ and $B_\pm(x,y)$ are {\it even} functions of $\Delta $.
Then it follows from \eqref{eq9b} that
\begin{equation}
D_\pm(x,y)=D_{\pm}(y,x), \quad B_\pm(x,y)=\pm B_\pm(y,x).\label{eq15}
\end{equation}
Since $B_-(x,y)$ is antisymmetric, we write
\begin{equation}
B_-(x,y)=(x-y)\bar B_-(x,y),\label{eq16}
\end{equation}
where $\bar B_-(x,y)$ is symmetric. Next we make a change of variables
\begin{equation}
u=xy,\quad v=x+y \label{eq17}
\end{equation}
and write
\begin{equation}
D_\pm(x,y)=d_\pm(u,v),\quad B_+(x,y)=b_+(u,v),\quad \bar B(x,y)=b_-(u,v). \label{eq18}
\end{equation}

As an example, let us consider solutions corresponding to $J^{(k)}=H^k$.
We denote the corresponding functions as $d^{(k)}_\pm(u,v)$ and $b_\pm^{(k)}(u,v)$.
A straightforward calculation gives
\begin{equation}
d_+^{(0)}=1,\quad d^{(0)}_-=0,\quad b_+^{(0)}=0,\quad b_-^{(0)}=0,\label{eq18a}
\end{equation}
\begin{equation}
d_+^{(1)}=u,\quad d^{(1)}_-=-1,\quad b_+^{(1)}={\ep}+gv,\quad b_-^{(1)}=0,\label{eq18b}
\end{equation}
\begin{align}
&d_+^{(2)}=u(u+1)+gv(2\ep+gv)+g^2+\Delta^2+{\ep^2},\quad d^{(2)}_-=-2u,\nonumber\\
&b_+^{(2)}=gv(2u+1)+2\ep u,\quad b_-^{(2)}=0,\label{eq19}
\end{align}
\begin{align}
d_+^{(3)}=&u(u^2+3u+3\Delta^2+1+{3\ep^2})+3\ep gv(2u+{1})+3g^2(u+1)v^2+g^2(3u+1),\nonumber\\
d^{(3)}_-=&-3u(u+1)-2\ep gv-g^2(1+v^2)-\Delta^2-{\ep^2},\nonumber\\
b_+^{(3)}=&{3\ep}u(u+1)+gv(1+{3\ep^2}+\Delta^2+3u(u+2))+
{3\ep}g^2(v^2+1)+\nonumber\\
&g^3v(v^2+3)+{\ep^3}+{\ep}\Delta^2,\nonumber\\ b_-^{(3)}=&g.\label{eq20}
\end{align}

We see that $b_-^{(k)}$, $k=1,2,3$ are very simple. For $k=4$
\begin{equation}
b_-^{(4)}=2g(1+2u).\label{eq21}
\end{equation}
For higher values of $k$, $b_-^{(k)}$ becomes a polynomial in both $u$ and $v$.

Now we can rewrite recurrence relations \eqref{eq9} as differential
equations for the functions
$A(x,y)$, $B(x,y)$, $C(x,y)$ and $D(x,y)$ defined as in \eqref{eq9c}.
Due to \eqref{eq12} only two of them are linearly independent
\begin{equation}
(\delta_x-\delta_y+2\Delta)B+{\ep}(D-A)+g((x+y)(D-A)+\partial_x D-\partial_y A)=0,\label{eq22}
\end{equation}
\begin{equation}
(\delta_x-\delta_y)A+{\ep}(C-B)+g((x+y)(C-B)+\partial_x C-\partial_y B)=0,\label{eq23}
\end{equation}
where we have denoted $\delta_x=x\partial_x$ and dropped the arguments $x$ and $y$ in all four functions.

Now we substitute \eqref{eq13} into (\ref{eq22})-(\ref{eq23}) and decouple the even and odd parts in $\Delta$ of both equations.
This gives four equations
\begin{align}
&(\delta_x-\delta_y)D_++g(\pa_x-\pa_y)B_+=0,\nonumber\\
&(\delta_x-\delta_y)D_-+2\ep B_-+g(2(x+y)B_-+(\pa_x+\pa_y)B_-)=0,\nonumber\\
&(\delta_x-\delta_y)B_++2\Delta^2B_-+g(\pa_x-\pa_y)D_+=0,\nonumber\\
&(\delta_x-\delta_y)B_-+2B_++2\ep D_-+g(2(x+y)D_-+(\pa_x+\pa_y)D_-)=0.\label{eq24}
\end{align}
Finally using \eqref{eq16} and \eqref{eq18} we rewrite them in symmetric variables $u$ and $v$
\begin{align}
&\pa_v d_+(u,v)-g\pa_u b_+(u,v)=0,\label{eq25}\\
&\pa_v d_-(u,v)+2(\ep+gv)b_-(u,v)+g(2\pa_v+v\pa_u)b_-(u,v)=0,\label{eq26}\\
&\pa_v b_+(u,v)-g\pa_u d_+(u,v)+2\Delta^2b_-(u,v)=0,\label{eq27}\\
&[(v^2-4u)\pa_v+v]b_-(u,v)+2b_+(u,v)
+[2\ep+g(2v+2\pa_v+v\pa_u)]d_-(u,v)=0.\label{eq28}
\end{align}
It is easy to check that (\ref{eq18a})-(\ref{eq20}) solve these four equations.

This is still a very complicated system of partial differential equations but much simpler than the original one.
To make further progress, let us assume that $b_-(u,v)$ is given by a finite series in $v$
\begin{equation}
b_-(u,v)=\sum_{k=0}^{M-1} f_k(u)v^k.\label{eq29}
\end{equation}
Here we set the upper limit to be $M-1$ to include the simple case  $b_-(u,v)=0$ when $M=0$.

Substituting \eqref{eq29} in \eqref{eq26} and integrating over $v$ we find $d_-(u,v)$ up to an arbitrary function
$h_1(u)$. Then from \eqref{eq28} we can find $b_+(u,v)$. Integrating \eqref{eq25} we find $d_+(u,v)$ up to
another arbitrary function $h_2(u)$. Substituting all functions in  \eqref{eq27}, we get
a finite series in $v$ with coefficients depending on $u$. Equating all coefficients at powers of $v$
to zero, we get a system of differential equations for $f_k(u)$ and $h_{1,2}(u)$ which can be solved
directly.

This program does not work (or is hard to implement)
for an infinite series in \eqref{eq29}, when we get an infinite system of
ODEs for functions $f_k(u)$. Another approach is to exclude functions $b_+(u,v)$, $d_+(u,v)$ and $d_-(u,v)$
from (\ref{eq25})-(\ref{eq28}) and derive the 2nd order partial differential equation for $b_-(u,v)$.
However, this complicated equation has polynomial coefficients in $u$ and $v$ and cannot be solved by any known
techniques. We shall not give it here.

We now turn to the specific details for the first few cases.

\noindent
(i) {\sl The case $M=0$}

In this case $b_-(u,v)=0$ and  integrating \eqref{eq25}, \eqref{eq26} and \eqref{eq28} gives
\begin{align}
&d_-(u,v)=h_1(u),\quad b_+(u,v)=-({\ep}+gv)h_1(u)-\frac{1}{2}gvh_1'(u),\\
&d_+(u,v)=h_2(u)-\frac{1}{2}gv(2\ep+gv)h_1'(u)-\frac{1}{4}g^2v^2h_1''(u). \label{eq30a}
\end{align}
Substituting \eqref{eq30a} into the last equation \eqref{eq27} we obtain a solution for $h_1(u)$ and $h_2(u)$
\begin{equation}
h_1(u)=\alpha e^{-2u}+\be_1+\be_2 u,\quad h_2(u)=-\be_1 u-\frac{\be_2}{2}u(u+1)+\be_3, \label{eq30}
\end{equation}
and a compatibility condition
\begin{equation}
\al\,\ep=0.\label{eq30b}
\end{equation}

The coefficients $\be_i$ correspond to ``gauge'' solutions (\ref{eq18a})-(\ref{eq19}).
The only nontrivial solution exists when $\ep=0$ and is given by
\begin{equation}
d_+(u,v)=b_+(u,v)=b_-(u,v)=0,\quad d_-(u,v)=e^{-2u}.\label{eq30c}
\end{equation}
Writing this solution back in the form \eqref{eq6} and dividing by $\Delta$ gives
\begin{equation}
J=\left(\begin{matrix}
\mathcal{P} &0 \\
0 & -\mathcal{P}\end{matrix}\right),\label{eq31}
\end{equation}
where $\mathcal{P}$ is the parity operator in the bosonic space
\begin{equation}
\mathcal{P}=:e^{-2a^\da a}:\label{eq31a}
\end{equation}
and we used the notation for normally ordered operators
\begin{equation}
:f(a^\da,a): \,\equiv \sum_{i,j=0}^\infty f_{i,j} (a^\da)^ia^j. \label{eq32}
\end{equation}
It is not difficult to check that
\begin{equation}
\mathcal{P}^2=1,\label{eq32a}
\end{equation}
and as a result
\begin{equation}
J^2=1. \label{eq32b}
\end{equation}

Finally, using the relation
\begin{equation}
:e^{-2a^\da a}:\>=e^{i\pi\,a^\da a}\label{eq33}
\end{equation}
we recover the well known  $\mathbb{Z}_2$ symmetry of the symmetric Rabi model
\begin{equation}
J=\sigma_z\otimes e^{i\pi\,a^\da a}.\label{eq34}
\end{equation}
The equation \eqref{eq33} follows from the identity
\begin{equation}
e^{\lambda a^\da a}=:e^{(e^\lambda-1)a^\da a}: \label{eq35}
\end{equation}
which can be proved by calculating the action of both sides of \eqref{eq35} on the Fock basis
\begin{equation}
|n\rangle=\frac{1}{\sqrt{n!}}(a^\da)^n|0\rangle. \label{eq36}
\end{equation}
Indeed,
\begin{equation}
e^{\lambda a^\da a} |n\rangle=e^{\lambda \mathcal{N}} |n\rangle=
e^{\lambda n} |n\rangle\label{eq36a}
\end{equation}
where $\mathcal{N}=a^\da a$ is the number operator. From the other side,
\begin{equation}
:e^{\mu a^\da a}:|n\rangle=\sum_{k=0}^\infty \frac{\mu^k}{k!}(a^\da)^k a^k|n\rangle=
\sum_{k=0}^n \frac{\mu^k}{k!}\frac{n!}{(n-k)!}|n\rangle=(1+\mu)^n|n\rangle,\label{eq36b}
\end{equation}
where we used
\begin{equation}
a^\da|n\rangle=\sqrt{n+1}|n+1\rangle,\quad a |n\rangle=\sqrt{n}|n-1\rangle.\label{eq36c}
\end{equation}
Comparing \eqref{eq36a} and \eqref{eq36b} gives \eqref{eq35}.

Applying the equivalence transformation $\eqref{eq4a}$ we can also rewrite
$J$ in the transformed basis
\begin{equation}
\widetilde{J}=UJU^{-1} = \sigma_x\otimes\mathcal{P}.\label{eq36d}
\end{equation}

\noindent
(ii) {\sl The case $M=1$}

Here we choose
\begin{equation}
b_-(u,v)=f(u).\label{eq37}
\end{equation}
Repeating the same procedure as in the previous section we find that there is a nontrivial integral
provided that
\begin{equation}
\ep=\pm\case{1}{2}. \label{eq38a}
\end{equation}
The corresponding solution has the form
\begin{align}
&d_+(u,v)=-\frac{\Delta^2}{g}e^{-2u},\quad
d_-(u,v)=2e^{-2u}(g-\ep v),\nonumber\\
&b_+(u,v)=0,\quad b_-(u,v)=e^{-2u}. \label{eq37a}
\end{align}
Dividing  $J$ by the overall factor $\Delta$ we obtain
\begin{equation}
J=\mathcal{P}\left(\begin{matrix}
2\ep(a^\da-a)+2g+\frac{\Delta}{g} &a^\da+a \\
-a^\da-a & 2\ep(a-a^\da)-2g+\frac{\Delta}{g} \end{matrix}\right),
\label{eq38}
\end{equation}
where $\mathcal{P}$  is the parity operator defined by \eqref{eq31a}.

It is an easy exercise to check that the operators \eqref{eq38} and
\eqref{eq3} commute when $\ep=\pm\frac{1}{2}$. We see that the operator $J$ is a product
of a parity operator $\mathcal{P}$ and a local matrix operator in the bosonic space.
In the case $\ep=0$ this local operator  was simply equal to $\sigma_z$.

Let us consider the case $\ep=\frac{1}{2}$.
In the transformed basis the operator $J$ becomes very simple
\begin{equation}
\widetilde{J}=UJU^{-1}=2\mathcal{P}\left(\begin{matrix}
\frac{\ds \Delta}{\ds 2g} &g-a \\
g+a^\da & \frac{\ds\Delta}{\ds 2g} \end{matrix}\right).\label{eq39}
\end{equation}

Finally, it is interesting to calculate $J^2$ to understand the spectrum of the operator $J$.
Using commutation relations
\begin{equation}
\mathcal{P}a^\da\mathcal{P}^{-1}=-a^\da,\quad \mathcal{P}a\mathcal{P}^{-1}=-a \label{eq40}
\end{equation}
one can easily obtain
\begin{equation}
J^2=4H+\lambda, \quad \lambda=4g^2+\frac{\Delta^2}{g^2}+2.\label{eq41}
\end{equation}
Therefore, for a given eigenvalue $E$ of the hamiltonian $H$ the operator $J$ has
one of the two possible eigenvalues $J(E)$, where
\begin{equation}
J(E)=\pm\sqrt{4E+\lambda}. \label{eq42}
\end{equation}
The whole Hilbert space will be again divided into two sectors with positive and negative
eigenvalues of $J$. However, this separation is ``dynamic'' in some sense because
it mixes degrees of freedom in bosonic and spin spaces.

\noindent
(iii) {\sl The case $M=2$}.

Here we calculate the operator $J$ for the case
\begin{equation}
b_-(u,v)=f_0(u)+f_1(u)v. \label{eq43}
\end{equation}
It only exists when $\ep=\pm1$ and
 we obtain after straightforward calculations
\begin{align}
&d_+(u,v)=2g e^{-2u}(2g\ep -v),
\quad b_+(u,v)=e^{-2u},\quad b_-(u,v)=\frac{2g^2}{\Delta^2}e^{-2u}(v-2g\ep ),\nonumber\\
&d_-(u,v)=e^{-2u}\left(\frac{4 g^2\ep}{\Delta^2}(u-g^2)+\frac{2g^2v}{\Delta^2}(2g-\ep v)-\ep\right).\label{eq44b}
\end{align}

For $\ep=1$ we obtain after a simple renormalization
\begin{equation}
J=\mathcal{P}\left(\begin{array}{cc}
2g^2((a^\da)^2+a^2)+2g(2g^2+\Delta)(a^\da-a)+(2g^2+\Delta)^2, \nonumber\\
2g^2(a^2-(a^\da)^2)-4g^3(a^\da+a)+\Delta,\end{array}\right.\label{eq44a}
\end{equation}
\begin{equation}
\left.\begin{array}{cc}
2g^2((a^\da)^2-a^2)+4g^3(a^\da+a)+\Delta \\
-2g^2((a^\da)^2+a^2)-2g(a^\da-a)(2g^2-\Delta)-(2g^2-\Delta)^2 \end{array}\right). \label{eq44}
\end{equation}
We notice that \eqref{eq38} and \eqref{eq44} satisfy \eqref{eq12} with the minus sign in the RHS.
This is because $J$ had a common factor $\Delta$ and we removed it for convenience.

Again the operator $J$ simplifies in the transformed basis
\begin{equation}
\widetilde{J}=\mathcal{P}\left(\begin{array}{cc}
(4g^2+1)\Delta+2g\Delta(a^\da-a) &\Delta^2+4g^2(g-a)^2\\
\Delta^2+4g^2(g+a^\da)^2 &(4g^2-1)\Delta+2g\Delta(a^\da-a)
\end{array}\right).\label{eq45}
\end{equation}
A straightforward calculation shows that $J^2$ is  a quadratic polynomial in $H$, namely
\begin{equation}
J^2=16g^4\,H^2+8g^2(4g^4+2g^2+\Delta^2)H+16g^6(g^2+1)+\Delta^2(8g^4+4g^2+1)+\Delta^4. \label{eq46}
\end{equation}
It is tempting to conjecture that  the off-diagonal entries of $\widetilde{J}$ depend separately on
$a$ and $a^\da$ as in \eqref{eq39} and \eqref{eq45}. However, for  $M>2$ this is not the case.
Apparently for any $\ep \in \frac12 \mathbb{Z}$ the integral $J$ is
a product of $\mathcal{P}$ and a matrix polynomial in $a^\da$, $a$ of degree $2\ep$.
Finally, we have calculated $J$ for $M=3,4,5$ and checked that
\begin{equation}
J^2=\sum_{i=0}^M \alpha_i(g,\Delta) H^i, \label{eq47}
\end{equation}
where coefficients $\alpha_i$ are polynomials in $g$ and $\Delta$. These coefficients quickly
become complicated for larger values of $M$. The fact that the symmetry operator $J$
is a square root of the polynomial in $H$ is very interesting and deserves further study.

It is not clear whether or not one can write a general formula for $J$, with $M=2\ep\in\mathbb{Z}_+$.
Finally, we notice that the authors of \cite{Gardas} argued the existence of the operator $J$
for generic values of $\ep$. However, their construction fails for $\ep \in \frac12 \mathbb{Z}$.
Moreover, their operator $J$ always satisfies the condition $J^2=1$ which is obviously
not the case in \eqref{eq47}. Since their result is mostly of an existential nature, we
can't compare it with our results.

\ack

The authors thank Daniel Braak and Masato Wakayama for instructive comments on the first version of this article.
In particular, Daniel Braak for pointing out that our results imply a $Z_2$ symmetry, and Masato Wakayama for remarks on
the self-adjoint nature of $J$.
This work has been supported by the Australian Research Council Discovery Projects DP170104934 and DP180101040.

\end{document}